\documentstyle[12pt]{article}
\newcommand{\be}{\begin{equation}}
\newcommand{\ee}{\end{equation}}
\newcommand{\ba}{\begin{eqnarray}}
\newcommand{\ea}{\end{eqnarray}}

\date{}
\begin{document}
\begin{flushright}
{\bf Preprint SPbU-IP-96-12\\
May 1996}\\
HEP-TH/9605007
\end{flushright}

\vspace{2cm}
\begin{center}
{\Large\bf Higher Order SUSY in Quantum Mechanics
and Integrability of Two-dimensional Hamiltonians}\footnote{This is a corrected
English version of the paper published in Russian
in: Problems in QFT and Statistical Physics, {\bf 13} (Eds. L.D.Faddeev,
A.G.Izergin and P.P.Kulish),
 Zapiski Nauch. Semin. POMI, 224 (1995) 68.
English transl. is to be published by AMS} \\
\vspace{0.5cm}

{\bf A. A. Andrianov}\footnote{E-mail: ANDRIANOV1@PHIM.NIIF.SPB.SU}, 
{\bf M. V. Ioffe}\footnote{E-mail: IOFFE@PHIM.NIIF.SPB.SU} and
{\bf D. N. Nishnianidze} \footnote{On leave of absence from
Kutaisi Polytechnic University, Georgia}\\
\vspace{0.5cm}

Department of Theoretical Physics,
University of Sankt-Petersburg,198904 Sankt-Petersburg, Russia.
\end{center}
\vspace{0.5cm}

\begin{flushright}
{\bf To Memory of V.N.Popov}
\end{flushright}

\vspace{0.5cm}

{\small 
\noindent
{\bf Abstract.} \qquad
The new method based on the SUSY algebra with supercharges of higher
order in derivatives is proposed to search for dynamical symmetry
operators in 2-dim quantum and classical systems. These symmetry 
operators arise when closing the SUSY algebra for a wide set of 
potentials. In some cases they are of 2-nd order in derivatives. 
The particular solutions are obtained also  for potentials accepting 
symmetry operators of 4-th order. The investigation of quasiclassical 
limit of the SUSY algebra yields new classical integrals of motion for a 
certain type of systems which are polynomials of 4-th order in momenta.
The general SUSY-inspired algorithm to construct classical systems
with additional integrals of motion is outlined.} 

\newpage

\section{Introduction}

The existence of dynamical symmetries in quantum systems
leads to their complete or partial integrability and therefore
gives the opportunity of a more detailed study of their spectral
properties.

We describe here the new method for 
searching the dynamical symmetry operators
based on the construction of the isospectral
Hamiltonians. The algebraic form of the isospectral transformations
in Quantum Mechanics is realized by the SUSY algebra 
\cite{Witten} - \cite{Lahiri}.

The standard one-dimensional Supersymmetrical Quantum Mechanics (SSQM)
is generated by supercharge operators $Q^{\pm},$ which form 
the SUSY algebra (together
with the Hamiltonian of supersystem $H )$:
\ba
\{Q^+, Q^-\} = H;\label{1}
\ea
\ba
[Q^{\pm}, H] = 0.\label{2}
\ea
This algebra is represented by $2 \times 2$ - matrix supercharges
\ba
Q^- = (Q^+)^{\dagger} = \left(\begin{array}{cc}
                           0&0\\
                           q^-&0\\
                            \end{array}\right);
\quad q^{\pm} = \mp \partial + W(x)
\ea
and by the superhamiltonian $H$ consisted of two Schr\"odinger operators:
\ba
H = \left(\begin{array}{cc}
      h^{(1)}&0\\
       0&h^{(2)}\\
      \end{array}\right) = \left(\begin{array}{cc}
                            q^+q^-&0\\
                               0&q^-q^+\\
                             \end{array}\right),
\quad h^{(i)} = -\partial^2 + V^{(i)}.
\ea
Eq.(\ref{2}), in terms of components, leads to the
intertwining relations for two Hamiltonians:
\be
h^{(1)}q^+ = q^+h^{(2)};\quad q^-h^{(1)} = h^{(2)}q^-.\label{3}
\ee
Therefore the spectra of $h^{(1)}$ and $h^{(2)}$ almost coincide
(up to zero - modes of the operators $q^{\pm}$) and their wave
functions are connected by $q^{\pm}$ mappings.

The extension of one-dimensional SSQM with supercharge
operators of higher order in derivatives was proposed in
\cite{AIS}, \cite{ACDI}. The related SUSY 
algebra turns out to be a  polynomial one, namely, the polynomial
of the Schr\"odinger
Hamiltonian appears in the right hand side of Eq.(\ref{1}).

Multidimensional generalizations of SSQM with supercharges
linear in derivative are known \cite{ABI}, \cite{ABIE}.
In the present paper the two-dimensional SSQM  generated by
supercharge operators of second order in derivatives
is investigated. In this case, in analogy to
1-dimensional SSQM we would expect 
that the corresponding SUSY algebra becomes
polynomial in a superhamiltonian. However, actually we find 
that the  SUSY algebra (\ref{1}) for
a wide set of potentials 
is closed by the diagonal operator of dynamical symmetry (see Sect.2).
It corresponds to the additional integral of motion for
the classical system (Sect.5). In some cases this symmetry operator
is of second order in derivatives (Sect.3) and the system admits the
$R$-separation of variables \cite{Miller}. But, in general
it is of 4-th order in derivatives. The particular solutions for
corresponding potentials and coefficient functions of
the symmetry operator are displayed in Sect.4. In Sect.5 the
quasiclassical limit of the superalgebra is studied and classical
integrals of motion of 4-th order were constructed (up to our
knowledge, some of them are new (compare to \cite{Per}).

\section{Two-dimensional Darboux transformations of se\-cond order
in derivatives}
\vspace{.5cm}
\hspace*{3ex}
Let us consider the intertwining relations (\ref{3}) retaining
the Planck constant $\hbar$ in the Hamiltonian and in the
supercharge
operators with components of general form:
\be
q^+ = (q^-)^{\dagger} = \hbar^2 g_{ik}{(\vec x)}\partial_i \partial_k +
 \hbar
C_i\partial_i + B. \label{4.1}
\ee
The form of metric $g_{ik}(\vec x)$ is determined by the intertwining
relations (\ref{3}):
\be
\partial_l g_{ik}{(\vec x)} +
\partial_i g_{lk}{(\vec x)} + \partial_k g_{il}{(\vec x)} =
0.\label{4.2}
\ee
Its solutions can be easily calculated:
\ba
&&g_{11}
= \widetilde{\alpha}x_2^2 + \widetilde a_1 x_2 + \widetilde b_1;
\label{4.3}\\ &&g_{22} = \widetilde{\alpha}x_1^2 + \widetilde
a_2 x_1 + \widetilde b_2; \label{4.4}\\ &&g_{12}
= -\frac{1}{2}(2\widetilde{\alpha}x_1 x_2 + \widetilde a_1 x_1 +
\widetilde a_2 x_2) + \widetilde b_3. \label{4.5}
\ea

Thus the senior in derivative part of supercharges
belongs to the $E(2)$ - universal enveloping algebra \cite{Miller}.
We distinguish four different classes in second derivatives:
\ba
&&q^{(1)+} = \gamma \triangle + C_i \partial_i + B;
\label{4.6}\\ &&q^{(2)+} = \alpha P_1^2 + \gamma \triangle +
C_i \partial_i + B; \label{4.7}\\
&&q^{(3)+} = \alpha \{J,P_1\} + \gamma \triangle +
C_i \partial_i + B; \label{4.8}\\ &&q^{(4)+} =
\alpha J^2 + \beta P_1^2 + \gamma \triangle + C_i
\partial_i + B, \label{4.9}
\ea
where $J$ and $\vec P$
are generators of rotations and translations, correspondingly,
and
$\alpha\not = 0$.

The coefficient functions of supercharge operator (\ref{4.1})
must satisfy differential equations:
\ba
&&\hbar\partial_i C_k + \hbar\partial_k C_i +
\hbar^2\triangle g_{ik} - (V^{(1)} - V^{(2)})g_{ik} = 0;
\label{4.10}\\ &&\hbar^2\triangle C_i + 2\hbar\partial_i B +
2\hbar g_{ik}\partial_k V^{(2)} - (V^{(1)} -
V^{(2)})C_i= 0;\label{4.11}\\ &&\hbar^2\triangle B +\hbar^2
g_{ik}\partial_k\partial_i V^{(2)} +\hbar C_i\partial_i V^{(2)}
- (V^{(1)} - V^{(2)}) B = 0.\label{4.12}
\ea
From Eqs.(\ref{3}) it is evident 
that the generalized superalgebra gives rise to 
the symmetry operator $\widetilde R$ for the Hamiltonian $H$:
\be
\{Q^+,Q^-\} =
\widetilde R;  \qquad [\widetilde R,H] = 0.\label{4.13}
\ee
This operator determines a higher-order dynamical symmetry which,
in general, can not be represented by the polynomial of
the Hamiltonian $H$ and of the second-order symmetry operator $R$.
As it will be shown in the  next Section the latter possibility exists
for the supercharges (\ref{4.6}) of the class 1 only. However in all
cases the closing of the SUSY algebra leads to the integrability of
the corresponding dynamical system.

The general solutions of nonlinear Eqs.
(\ref{4.10}) - (\ref{4.12}) for all four classes of the metric are not
known. But in some cases the particular solutions can be found in
the analytical form.

\section{Particular solutions of class 1: dynamical sym\-met\-ries of
second order}
\vspace{.5cm}
\hspace{3ex}
Let us solve the system of differential Eqs.
(\ref{4.10}) - (\ref{4.12}) for the supercharges with the diagonal
metric $g = diag(-1,-1)$ $(\hbar  = 1):$
\ba
q^+ = -\triangle + C_i \partial_i + B ; \quad  q^- =
(q^+)^{\dagger}.
\label{5.1}
\ea
From Eqs.(\ref{4.10}), (\ref{4.11}) it follows that
\ba
&&C^2 \equiv (C_1 + iC_2)^2 = \alpha z^2 + 8\beta z +
\gamma;\quad z \equiv x_1 + ix_2;\label{5.2}\\
&&V^{(2)}-B = \frac{1}{4}\alpha\mid z\mid^2
+ z\bar \beta + \bar z\beta + \frac{1}{4}\mid C\mid^2 - \eta;
\label{5.3}\\&&V^{(2)}-V^{(1)} = \partial_z C + \partial_{\bar
z}\bar C,\label{5.4}
\ea
where
$\alpha$, $\eta$
are real constants and
$\beta $, $\gamma $
may be complex ones.
Eq. (\ref{4.12})
can be written in the form:
\ba
(C\partial_z + \bar
C\partial_{\bar z})(B\mid C\mid^2) = G\mid C\mid^2.\label{5.5}
\ea
where
\be
G = \alpha +
(\partial_zC)(\partial_{\bar z}\bar C) - \frac{\alpha}{2}(\bar
z C + z\bar C) - 2(\bar \beta C + \beta \bar C ).\nonumber
\ee
To solve (\ref{5.5}) it is useful to change the variables
$\bar z$  to $\tau_1$, $\tau_2$:
\be
\tau_1 =
\int\frac{dz}{C} + \int\frac{d\bar z}{\bar C}; \quad i\tau_2 =
\int\frac{dz}{C} - \int\frac{d\bar z}{\bar C}.\label{5.6}
\ee
Then
\be
B = \frac{1}{2\mid C\mid^2}\int G\mid C\mid^2d\tau_1 +
\frac{F(\tau_2)}{\mid C\mid^2}, \label{5.7}
\ee
where $F(\tau_2)$ is an arbitrary real function. Thus the irreducible
SUSY algebra of second order in derivatives is realized for  a broad
variety of potentials.

For the class 1 the superalgebra (\ref{4.13})
gives rise to the symmetry operator
$R$ of second order in derivatives:
\be
\widetilde R = H^2 + R + 2\eta H, \label{5.8}
\ee
where $R$ is diagonal operator:
$$
R = \left( \begin{array}{cc}
       R^{(1)} & 0 \\
       0   & R^{(2)} \\
      \end{array} \right).
$$
Its components
\ba
R^{(1)}&=&2\bigl(\alpha \mid z\mid^2 + 4(\bar \beta z + \beta
\bar z )\bigr)\partial_z \partial_{\bar z} - C^2\partial_z^2 -
\bar C ^2 \partial_{\bar z}^2 -
C (\partial_z C)\partial_z - \nonumber\\
&&\bar C (\partial_{\bar z} \bar C)\partial_{\bar z} +
(C\partial_z + \bar C\partial_{\bar z})(B + V^{(1)})
+ B^2 - V^{(1)2} -2\eta V^{(1)}; \label{5.9}
\ea
\ba
R^{(2)}&=&2\bigl(\alpha \mid z\mid^2 +
4(\bar \beta z + \beta \bar z )\bigr)\partial_z \partial_{\bar
z} - C^2\partial_z^2 - \bar C ^2\partial_{\bar z}^2
-C(\partial_z C)\partial_z -\nonumber\\&&\bar
C (\partial_{\bar z} \bar C )\partial_{\bar z} - (C\partial_z +
\bar C \partial_{\bar z})(B - V^{(2)}) +
B^2 - V^{(2)2} -2\eta V^{(2)} \label{5.10}
\ea
are symmetry operators for the Hamiltonians
$h^{(1)}$, $h^{(2)}$,
respectively.

Let us find explicit expressions for the operators $H, Q^{\pm}, R$.
They 
depend on the values of $\alpha$ and $\beta$ in Eq.(\ref{5.2}).

i) If  $\alpha= 0$, $\beta\not= 0$, the constant $\gamma$
in (\ref{5.2}) can be excluded by the translation
 $z\rightarrow z - (\gamma/8\beta)$.  Then the variables (\ref{5.6})
read
$$ \tau_1 = \sqrt{\frac{z}{2\beta}} +
\sqrt{\frac{\bar z}{2\bar \beta}}; \quad \tau_2 = i\Bigl(\sqrt{\frac{\bar
z}{2\bar \beta}} - \sqrt{\frac{z}{2\beta}}\Bigr) $$
and they are  related  to the conventional \cite{Miller} parabolic coordinates.
It follows from Eq.(\ref{5.7}) that
$$ B =\frac{2\tau_1 - \mid \beta\mid^2\tau_1^4 - 2\mid \beta
\mid^2 \tau_1^2 \tau_2^2 + F(\tau_2)}{\tau_1^2 + \tau_2^2}. $$
Correspondingly, from Eqs.(\ref{5.3}), (\ref{5.4}) we find the potentials $V^{(1,2)}$:
\ba
V^{(i)}= \frac{(-1)^i 2\tau_1 +
\mid \beta \mid^2 \tau_1^4 + F(\tau_2)}{\tau_1^2 + \tau_2^2} -
\eta; \label{5.11}
\ea
and from Eqs.(\ref{5.9}), (\ref{5.10}) --
the components of the symmetry operator $R$:
$$
R^{(i)}= \frac{4}{\tau_1^2 +
\tau_2^2}(\tau_1^2\partial_{\tau_2}^2 -
\tau_2^2\partial_{\tau_1}^2)+\frac{4\mid \beta
\mid^2 }{\tau_1^2 + \tau_2^2}\biggl[((-1)^i 2\tau_1 + \mid \beta
\mid^2 \tau_1^4)\tau_2^2 - \tau_1^2 F(\tau_2)\biggr] -
\eta^2.
$$
In terms of variables $\tau_1$, $\tau_2$ the Laplacian
$$ \triangle = \frac{1}{\mid \beta \mid^2(\tau_1^2 +
\tau_2^2)}(\partial_{\tau_1}^2 + \partial_{\tau_2}^2)$$
contains the same multiplier as in Eqs.(\ref{5.11}).
Thus the spectral problem for both Hamiltonians, $h\Psi= E\Psi$,
can be solved by the
$ R $-separation of variables \cite{Miller}. Namely, the corresponding
eigenfunctions (of $h^{(1)}$ and $h^{(2)}$) with  a given energy $ E $
can be expanded into the sum
\be
\Psi = \sum_n{\nu_n \phi_{1n}(\tau_1)
\phi_{2n}(\tau_2)}, \label{5.12}
\ee
where $\nu_n$ are constants, and $\phi_{1n}(\tau_1)$,
$\phi_{2n}(\tau_2)$ are solutions of 1-dimensional equations
\ba
&& -\phi_{1n}''(\tau_1) + \mid \beta \mid^2[-(E+\eta)\tau_1^2
\mp 2\tau_1 + \mid\beta\mid^2\tau_1^4] \phi_{1n}(\tau_1) =
\frac{\lambda_n^{\pm}}{4}\phi_{1n}(\tau_1); \nonumber\\
&&-\phi_{2n}''(\tau_2) + \mid\beta\mid^2[-(E+\eta)\tau_2^2 +
F(\tau_2)] \phi_{2n} =
-\frac{\lambda_n^{\pm}}{4}\phi_{2n}(\tau_2).  \label{5.13}
\ea
The upper (lower) sign in Eq.(\ref{5.13}) corresponds to
$h^{(1)}(h^{(2)})$, and $\lambda_n^{\pm}$ are constants of separation,
which play  role of the spectral parameters for the symmetry
operators $R^{(1)}$ ($R^{(2)}$) respectively:
$$R\phi_{1n}(\tau_1)\phi_{2n}(\tau_2) = (\lambda_n -
\eta^2)\phi_{1n}(\tau_1)\phi_{2n}(\tau_2).$$

Depending on the properties of the function $F(\tau_2)$ the
degeneracy of  energy levels, i.e.  the dimension of the space  $\{\lambda_i\}$ 
for a given $E$ can be finite or even infinite.

 ii) Let us describe the second case $\beta= 0$, $\alpha > 0,$
when the suitable coordinates,
\ba
&&\tau_1 = \frac{1}{\sqrt\alpha}\ln\Biggl[\biggl(z + \sqrt{z^2 +
\frac{\gamma}{\alpha}}\biggr)\biggl(\bar z + \sqrt{\bar z ^2 +
\frac{\bar \gamma}{\alpha}}\biggr)\Biggr];\label{5.14}\\
&&\tau_2 = - \frac{i}{\sqrt\alpha}\ln\frac{z + \sqrt{z^2 +
\frac{\gamma}{\alpha}}}{\bar z + \sqrt{\bar z ^2 +
\frac{\bar \gamma}{\alpha}}},\label{5.15}
\ea
are connected to the conventional elliptic ones.
In full analogy with the case i) one can obtain that:
\ba
B&=&\frac{1}{2(f_1 + f_2)}(2\partial_{\tau_1}f_1 -
\frac{1}{2}f_1^2 - f_1f_2 + F(\tau_2));\label{5.16}\\
V^{(i)}&=&\frac{1}{2(f_1 + f_2)}((-1)^i
2\partial_{\tau_1}f_1 + \frac{1}{2}f_1^2 + F(\tau_2)) - \eta;\label{5.17}\\
R^{(i)}&=&\frac{4}{f_1 + f_2}(f_1\partial_{\tau_2}^2 -
f_2\partial_{\tau_1}^2) +\nonumber\\
&&\frac{f_2((-1)^i 2\partial_{\tau_1}f_1 + f_1^2/2)
- f_1 F(\tau_2)} {f_1 + f_2} - \eta^2,\label{5.18}
\ea
where
\ba &&f_1 =
\frac{1}{4}\biggl(\alpha \exp(\sqrt{\alpha}\cdot \tau_1) +
\frac{\mid\gamma\mid^2}{\alpha} \exp(-\sqrt{\alpha}\cdot\tau_1)
\biggr);\label{5.19}\\ &&f_2 = \frac{1}{4}\biggl(\bar \gamma
\exp(i\sqrt{\alpha}\tau_2) + \gamma
\exp(-i\sqrt{\alpha}\cdot\tau_2)\biggr).\label{5.20}
\ea
In terms of the coordinates $\tau_1$, $\tau_2$
the solution for  eigenfunctions can be again decomposed into  the sum
(\ref{5.12}), where now the functions   $\phi_{1n}(\tau_1);\,
\phi_{2n}(\tau_2)$ satisfy the equations:
$$-\phi_{1n}''(\tau_1) +
\frac{1}{4}\Biggl(\mp \partial_{\tau_1}f_1 + \frac{1}{2}f_1^2 -
 (E + \eta)f_1\Biggr)\phi_{1n}(\tau_1) =
 \frac{\lambda_n^{\pm}}{4}\phi_{1n}(\tau_1);$$
$$-\phi_{2n}''(\tau_2) + \frac{1}{4}\Biggl(\frac{F(\tau_2)}{2} -
 (E + \eta)f_2\Biggr)\phi_{2n}(\tau_2) =
-\frac{\lambda_n^{\pm}}{4}\phi_{2n}(\tau_2)
$$
and $\lambda_n^{\pm}$ are eigenvalues of the symmetry operator
(\ref{5.18}).

The analysis of the case $\beta= 0$, $\alpha < 0$ is similar
and its result can be formulated as follows.
It is necessary to replace real ($\tau_1,\tau_2$) in Eqs.(\ref{5.14}),
(\ref{5.15}) to imaginary
($-i\tau_2,-i\tau_1$) and, respectively,
($f_1(\tau_1),f_2(\tau_2)$ ) in Eqs.(\ref{5.19}), (\ref{5.20}) to
($-f_2(\tau_2),-f_1(\tau_1)$). Then  the relations for $B,
V^{(1,2)}, R^{(1,2)}$ are given by Eqs.(\ref{5.16})-(\ref{5.18}).

Note that for specially chosen $\alpha > 0$, $\beta= \gamma= 0 $
the supercharge is factorized into the product of two conventional
superoperators \cite{ABI} that corresponds to the separation of variables
in the polar coordinates. Thus the supersymmetry, i.e. the intertwinning
relations between two Hamiltonians, leads inevitably to the hidden
dynamical symmetry realized by the operator
$ R $ and moreover to the
$ R $ - separation of variables in the case when  a supercharge operator
contains senior derivatives in the form of Laplacian.

\section{Particular solutions of class 2: dynamical sym\-met\-ries of fourth
order}
\vspace{.5cm}
{\bf I.}\quad
Let us consider the intertwinning relations (\ref{3}) for the supercharge
(\ref{4.1}) with the metric $g = diag(1,-1)$:
\be
q^+ = \hbar^2 (\partial_1^2 - \partial_2^2)
+ \hbar C_k \partial_k + B.\label{6.1}
\ee
From Eqs.(\ref{4.10})-(\ref{4.12}) we obtain, that:
\ba
&& C_{\pm} \equiv C_1 \mp C_2 =
C_{\pm}(x_{\pm});\label{6.2}\\
&&B = \frac{1}{4}(C_+ C_- + F_1(x_+ + x_-) +
F_2(x_+ - x_-));\label{6.4}\\ &&\partial_-(C_- F) =
-\partial_+(C_+ F);\quad F = F_1(x_+ + x_-) + F_2(x_+ -
x_-),\label{6.5}
\ea
and,  respectively, for potentials:
\ba
V^{(1,2)}& =& \pm\frac{\hbar}{2}(C_+' + C_-') + \frac{1}{8}(C_+^2 + C_-^2=
) +
\nonumber\\&&\frac{1}{4}(F_2(x_+ -x_-) - F_1(x_+ + x_-)) + const,\label{VL}
\ea
where
$x_{\pm} \equiv x_1 \pm x_2$; \quad
$\partial_{\pm} = \partial  / \partial x_{\pm}$.

Eqs.(\ref{6.5}) can be solved in  certain cases:

1) Let $C_- = 0,$ then
\ba
C_+(x_+)& =& \frac{1}{\delta_1 \exp(\sqrt{\lambda}
\cdot x_+) + \delta_2 \exp(-\sqrt{\lambda}\cdot
x_+)};\label{6.6}\\
F_1& =& \sigma_1 \delta_1
\exp(\sqrt{\lambda}\cdot (x_- + x_+)) + \sigma_2 \delta_2
\exp(-\sqrt{\lambda}\cdot (x_- + x_+));\label{6.7}\\ F_2& =&
\sigma_1 \delta_2 \exp(\sqrt{\lambda}\cdot (x_+ - x_-)) +
\sigma_2 \delta_1 \exp(-\sqrt{\lambda}\cdot (x_+ - x_-)).
\label{6.8}
\ea
Here and in  what follows the Greek letters stand for constants. Depending
on the sign $\lambda$ the latter ones may be real or complex.

2) Let the function $ F $ allow the factorization:
$F = F_+(x_+)\cdot F_-(x_-)$.
Then from Eq.(\ref{6.5}) we obtain that:
\be
C_{\pm} = \frac{\nu_{\pm}}{F_{\pm}} \pm
\frac{\gamma}{F_{\pm}}\int
\limits^{x_{\pm}}F_{\pm}dx'_{\pm},\label{6.10}
\ee
and there appear two possibilities:
\ba
&&a)\quad F_{\pm}(x_{\pm}) =
\epsilon_{\pm}x_{\pm},\label{6.11}\\&&b)\quad F_{\pm} =
\sigma_{\pm} \exp(\sqrt{\lambda}\cdot x_{\pm}) + \delta_{\pm}
\exp(-\sqrt{\lambda}\cdot x_{\pm}).\label{6.12}
\ea

Note that  the constant was
omitted in the r.h.s. of Eq.(\ref{6.11}) since the constant
solutions $F_{\pm}$ can be deduced from Eq.(\ref{6.12})
with $\lambda = 0$.

In the case $a$) we obtain:
$$
F_1(x_- + x_+) = \frac{\epsilon_+ \epsilon_-}{4}(x_+ +
x_-)^2;\quad F_2(x_+ - x_-) = -\frac{\epsilon_+
\epsilon_-}{4}(x_+ - x_-)^2;
$$
and in the case $b$):
\ba
F_1(x_+ + x_-) = \sigma_+\sigma_- \exp(\sqrt{\lambda}\cdot (x_+
+ x_-)) + \delta_+\delta_- \exp(-\sqrt{\lambda}\cdot (x_+ +
x_-)); \label{f1}\\ F_2(x_+ - x_-) = \sigma_+\delta_-
\exp(\sqrt{\lambda}\cdot (x_+ - x_-)) + \sigma_-\delta_+
\exp(-\sqrt{\lambda}\cdot (x_+ - x_-))\label{f2}
\ea
In the  cases 1), 2$a$) and 2$b$) potentials can be
found  by means of Eq.(\ref{VL}).

For this metric the components of the symmetry operator $\widetilde R$
(\ref{4.13}) are given by:
\ba
\widetilde R^{(1,2)}&=&16\hbar^2 \partial_+^2\partial_-^2 +
( \mp 4\hbar\partial_- C_- - C_-^2 )\hbar^2\partial_+^2 +
( \mp 4\hbar\partial_+ C_+ - C_+^2 )\hbar^2\partial_-^2 +\nonumber\\
&&2 ( 4B - C_+C_- )\hbar^2\partial_+\partial_- +
( 4\hbar\partial_-B - \hbar C_+\partial_-C_- \mp BC_- )\hbar\partial_+
+\nonumber\\ &&( 4 \hbar\partial_+B - \hbar C_- \partial_+C_+ \mp BC_+ )
\hbar\partial_- +\nonumber\\
&&B^2 + \hbar ( C_-\partial_+B + C_+\partial_-B ) +
4\hbar^2\partial_+\partial_-B.\label{sym1}
\ea

\noindent
{\bf ${\hbox{I}}\!{\hbox{I}}$.}\quad
Let us examine now the subclass of degenerate metric $g= diag(1,0),$
when the supercharges take the form,
\be
q^+ = \hbar^2\partial_1^2
+ \hbar C_k\partial_k + B.\label{q}
\ee
It follows from Eqs.(\ref{4.10})-(\ref{4.12}) that:
\ba
&&C_1 = -x_2 F_1'
+ G_1; \quad C_2 = F_1;\label{k0}\\ &&V^{(1)} = \hbar (2G_1' -
x_2 F_1'') + \frac{1}{4}x_2^2 (F_1^2)'' - x_2 (F_1G_1)' +
\nonumber\\ &&K_1(x_1) + K_2(x_2);\label{k1}\\ &&V^{(2)} =
\hbar x_2 F_1'' + \frac{1}{4}x_2^2 (F_1^2)'' - x_2 (F_1G_1)' +
K_1(x_1) + K_2(x_2);\label{k2}\\ &&B = -\frac{\hbar}{2}(G_1' +
x_2 F_1'') + \frac{1}{2}G_1^2 - \frac{1}{2}x_2^2F_1F_1'' + x_2
F_1G_1' - K_1(x_1),\label{k3}
\ea
where $F_1, G_1, K_1$ are arbitrary real functions of variable
$x_1$, and
$K_2(x_2)$ depends on $x_2$ only.

After substitution of Eqs. (\ref{k0}) - (\ref{k3})
into Eq.(\ref{4.12}) we arrive to the equation for the coefficient
functions:
\ba
&&-\frac{\hbar^2}{2}G_1''' + \frac{\hbar}{2}((G_1^2)'' +
2G_1'^2) + G_1K_1' + 2G_1'K_1 - F_1(F_1G_1)' -
G_1'G_1^2 + \nonumber\\&&x_2\biggl[
\frac{\hbar^2}{2}F_1^{(IV)} - \hbar(F_1'G_1'' + 2G_1'F_1'') -
G_1(2G_1'F_1' + G_1''F_1) -\nonumber\\&&F_1'K_1' + \frac{1}{2}
F_1(F_1^2)'' - 2G_1'^2F_1 - 2F_1''W_1 +
\biggr] + \nonumber\\&&x_2^2\biggl[
\frac{1}{4}G_1(F_1^2)''' + F_1'(F_1G_1)'' +
3G_1'F_1F_1''\biggr] - \nonumber\\&&x_2^3
\biggl[\frac{1}{4}F_1'(F_1^2)''' + F_1F_1''^2 \biggr] +
F_1K_2'(x_2) = 0.\label{k4}
\ea
It means that $K_2(x_2)$ is the polynomial of $x_2$ with  constant
coefficients:
\be
K_2(x_2) = m_0  - m_1 x_2 - \frac{1}{2}m_2 x_2^2 - \frac{1}{3}m_3 x_2^3
+ \frac{1}{4}m_4 x_2^4.\label{k9}
\ee
Therefore Eq.(\ref{k4}) is equivalent to four equations:
\ba
&&-\frac{\hbar^2}{2}G_1''' + \frac{\hbar}{2}((G_1^2)'' +
2G_1'^2) + G_1K_1' + \nonumber\\&&2G_1'K_1 - F_1(F_1G_1)' -
G_1'G_1^2 = m_1F_1;\label{k5}\\
&&\frac{\hbar^2}{2}F_1^{(IV)} -
\hbar(F_1'G_1'' + 2G_1'F_1'') - G_1(2G_1'F_1' + G_1''F_1)
-\nonumber\\&&F_1'K_1' + \frac{1}{2} F_1(F_1^2)'' - 2G_1'^2F_1 -
2F_1''K_1 = m_2F_1;\label{k6}\\
&&\frac{1}{4}G_1(F_1^2)''' + F_1'(F_1G_1)'' + 3G_1'F_1F_1'' =
m_3F_1;\label{k7}\\
&&\frac{1}{4}F_1'(F_1^2)''' + F_1F_1''^2 = m_4F_1. \label{k8}
\ea
It  happens to be possible to solve the system of differential Eqs.
(\ref{k5}) - (\ref{k8}) in particular cases. These solutions
will be presented for the quasiclassical limit
$\hbar \to 0$ in the next Sec.

\section{Quasiclassical limit and  integrable systems of \mbox{class 2}}
\vspace{.5cm}
\hspace*{3ex}
The quantum dynamical symmetries which were found by the intertwinning
method in previous Sections have the natural analogs,  integrals
of motion -- in the corresponding classical systems. For the class 1 these
integrals of motion of  2-nd order are known \cite{Per}.

Let us describe certain two-dimensional classical systems of class 2 with
additional integrals of motion of the 4-th order in derivatives.
Systems of this type are discussed in the book \cite{Per}, where
the list of known particular solutions for the coefficient function of
integrals of motion was presented.

\noindent {\bf I.}\quad
Let us define momenta $p_{\pm} =
-i\hbar\partial_{\pm}$ and take the limit $\hbar \to 0$.  Then
for the  Lorentz metric we derive from Eqs.(\ref{VL}),
(\ref{sym1}),
that the classical Hamiltonian $$h_{cl} = 2 ( p_+^2 + p_-^2 ) +
V_{cl},$$ where
\ba
V_{cl} = \frac{1}{8}( C_+^2 + C_-^2 ) +
\frac{1}{4}( F_2(x_+ - x_-) - F_1(x_+ + x_-))\label{7.1}
\ea
reveals the additional integral of motion
\ba
I = 16p_+^2p_-^2 + C_+^2p_-^2 + C_-^2p_+^2 - 2 ( F_1 +F_2)p_+p_- +
B^2.\label{7.2}
\ea
The functions $C_{\pm}, F_i, B$ were obtained for certain cases in Sect. 4.

\noindent
{\bf ${\hbox{I}}\!{\hbox{I}}$.}\quad
Let us investigate  what kind of integrable systems can be obtained from
the SSQM with degenerate metric (see Sect.4). At first, let us
define classical functions of supercharge from Eq.(\ref{q}):
\ba
q_{cl}^+ = -p_1^2 +i C_k p_k + B;\quad
q_{cl}^- = - p_1^2 - i C_k p_k + B.\label{7.3}
\ea
Thus the Hamiltonian $$h_{cl} = p_k^2 + V_{cl},$$ where
\be
V_{cl} = \frac{1}{4} x_2^2 (F_1^2)'' - x_2 (F_1G_1)' + K_1(x_1)
+ K_2(x_2), \label{7.4}
\ee
has the integral of motion
\ba
I \equiv q_{cl}^+q_{cl}^- = p_1^4 + ( C_1^2 - 2B ) p_1^2 + C_2^2
p_2^2 + 2C_1C_2 p_1p_2 + B^2.\label{7.5}
\ea
The functions $C_k$ are defined according to Eq.(\ref{k0}) and
other functions satisfy the system of equations (\ref{k5}) - (\ref{k8})
with $\hbar = 0.$

Let us display the non-trivial solutions which have been found by now
($k_i$ are constants):

1)\qquad $F_1 = k_1\not= 0.$
\noindent
Then the constants $m_3= m_4= 0,$ and  the function $G_1(x_1)$
satisfies the equation
\ba
\int \frac{G_1^2 dG_1}{\sqrt{k - \frac{1}{2}m_2G_1^4}} =
x_1,\label{G}
\ea
in turn, the function $K_1(x_1)$ can be written in terms of $G_1(x_1):$
\ba
K_1(x_1) = k_2G_1^{-2} + \frac{1}{4}G_1^2 + k_1m_1 G_1^{-2}\int
G_1 dx'.\label{K}
\ea
For nonzero values of constants $k$ and $m_2$
the integral can be represented as a sum of elliptic  functions of
1-st and 2-nd genus. However  in two limits 
$G_1(x_1)$ can be an elementary function when
a)\quad $k > 0,$ \quad $m_2 = 0$; \qquad b)\quad $k = 0,$ \quad $m_
2 < 0.$

In the case a), after the redefinition of constants $k_1,$ $k_2$
and an appropriate shift of $x_2$,
we find from Eqs.(\ref{7.4}), (\ref{7.5}), (\ref{K}),
taking into account (\ref{G}), that
\ba
&&V_{cl} = - \frac{k_1k_2}{3} x_2x_1^{-2/3} +
\frac{1}{4}\bigl[k_2^2 + \frac{3k_1m_1}{k_2}\bigr] x_1^{2/3} -
m_1 x_2 + const;\label{PV}\\ &&I = p_1^4 + \biggl[
\frac{1}{2}(k_2^2 + \frac{3k_1m_1}{k_2})x_1^{2/3} -
\frac{2k_1k_2}{3} x_2x_1^{-2/3} + k_1^2 \biggr] p_1^2 + k_1^2
p_2^2 +\nonumber\\ &&2k_1k_2x_1^{1/3}p_1p_2 + \biggl[
\frac{k_1k_2}{3} x_2x_1^{-2/3} + \frac{1}{4}(k_2^2 -
\frac{3k_1m_1}{k_2}) x_1^{2/3} -
\frac{k_1^2}{2}\biggr]^2.
\ea

In the case b) and other solvable cases such as
2) $G_1 = 0,$\quad $F_1= k_1x_1$ or $F_1= k_1x_1^2;$
3) $F_1 = k_1x_1^2$,\quad $G_1 = k_2x_1;$
4) $F_1 = k_1x_1$,\quad $G_1 = k_2;$
5) $F_1 = 0,$
the potentials  are obtained  as  cited in  \cite{Per} (i.e. the potentials
with separable variables),
that is why they are not displayed here.

When comparing the potentials (\ref{7.1}) (with the coefficient functions
$C_{\pm}, F_1, F_2$ in the cases 1) and 2$b$) of Sec. 4)
 with the list in \cite{Per} we find that they are new and posess the dynamical
symmetries of 4-th order.

In conclusion let us formulate the recipe that we have established 
for the construction
of integrals of motion in  Classical Mechanics. For a given
classical Hamiltonian $h_{cl}^{(1)}$ we look for the complex function
$q_{cl}^+(\vec x, \vec p) = (q_{cl}^-)^{\dagger},$ polynomial
in momenta and such that its Poisson brackets
with $h_{cl}^{(1)}$  have the form:
\ba
\{q_{cl}^-, h_{cl}^{(1)}\} =
-i f(\vec x, \vec p)q_{cl}^-;\quad \{q_{cl}^+, h_{cl}^{(1)}\} =
i f(\vec x, \vec p)q_{cl}^+,
\ea
where $f(\vec x, \vec p)$ is an arbitrary real function. Ii is evident that
then the classical factorizable integral of motion
$I = q_{cl}^+q_{cl}^-$ exists.

This paper was supported by the RFBI grant 
 No.96-01-00535 and by the GRACENAS grant No.95-0-6.4-49.


\vspace{.5cm}
\begin{thebibliography}{}
\bibitem{Witten} Witten. E., Nucl. Phys. B 188 (1981) 513;
B 202 (1982) 253.
\bibitem{Gend}
Gendenstein L.E., Krive I.V., Usp. Fizich. Nauk (Russ. UFN) 146 (1985) 553 .
\bibitem{Lahiri}  Lahiri A., Roy P.K. and
Bagghi B., Int. J.  Mod. Phys. A 5 (1990) 1383.
\bibitem{AIS} Andrianov A.A., Ioffe M.V. and Spiridonov V.P.,
Phys. Lett. A 174 (1993) 273.
\bibitem{ACDI}
Andrianov A.A., Cannata F., Dedonder J.-P. and Ioffe M.V.,
Int. J. Mod. Phys. A10 (1995) 2683.
\bibitem{ABI}
Andrianov A.A., Borisov N.V. and Ioffe M.V., Teor.Mat.Fiz. 61 (1984)
183;  Phys. Lett. A 105 (1984) 19.
\bibitem{ABIE}
Andrianov A.A., Borisov N.V., Ioffe M.V. and Eides M.I.,
Teor.Mat.Fiz. 61 (1984) 17; Phys.  Lett. A 109 (1985) 143.
\bibitem{Miller}
Miller W.,Jr., "Symmetry and Separation of Variables".
Addison-Wesley Publishing Company, London, 1977.
\bibitem{Per}
Perelomov A.M.,  "Integrable systems of classical mechanics and Lie
algebras"(in Russian). Moscow, Nauka, 1990.
\end{thebibliography}
\end{document}